\documentclass[twocolumn,showpacs,preprintnumbers,amsmath,amssymb,superscriptaddress]{revtex4}

\allowdisplaybreaks
\allowdisplaybreaks[4]
\usepackage{CJK}
\usepackage{graphicx}
\usepackage{dcolumn}
\usepackage{bm}
\usepackage[dvipdfm,
            pdfstartview=FitH,
            CJKbookmarks=true,
            bookmarksnumbered=true,
            bookmarksopen=true,
            colorlinks,
            pdfborder=001,
            linkcolor=blue,
            anchorcolor=blue,
            citecolor=blue
            ]{hyperref}

\begin{document}

\begin{CJK}{GBK}{song}

\title{Wobbling geometry in simple triaxial rotor}

\author{W. X. Shi}
\affiliation{State Key Laboratory of Nuclear Physics and Technology,
             School of Physics, Peking University, Beijing 100871, China}%
\affiliation{Yuanpei College, Peking University, Beijing 100871, China}

\author{Q. B. Chen}\email{qbchen@pku.edu.cn}
\affiliation{State Key Laboratory of Nuclear Physics and Technology,
             School of Physics, Peking University, Beijing 100871, China}%

\date{\today}

\begin{abstract}
  The spectroscopy properties and angular momentum geometry for the wobbling motion of
  a simple triaxial rotor are investigated within the triaxial rotor model up to spin $I=40\hbar$.
  The obtained exact solutions of energy spectra and reduced quadrupole transition probabilities
  are compared to the approximate analytic solutions by harmonic approximation formula and
  Holstein-Primakoff formula. It is found that the low lying wobbling bands can be well
  described by the analytic formulas. The evolution of the angular momentum geometry as well as the
  $K$-distribution with respect to the rotation and the wobbling phonon excitation are studied in
  detail. It is demonstrated that with the increasing of wobbling phonon number, the triaxial rotor
  changes its wobbling motions along the axis with the largest moment of inertia to the axis with the
  smallest moment of inertia. In this process, a specific evolutionary track that can be used to
  depict the motion of a triaxial rotating nuclei is proposed.
\end{abstract}

\pacs{21.60.Ev, 
21.10.Re, 
23.20.Lv  
} 

\maketitle


\section{Introduction}
\label{intro}

The concept of wobbling motion was originally proposed more than forty years ago by
Bohr and Mottelson when they investigated the rotational modes for a triaxial
nucleus using triaxial rotor model (TRM)~\cite{Bohr1975}. In the pioneering
work~\cite{Bohr1975}, they pointed out that the rotational angular momentum for a
triaxial nucleus is not aligned along any of the body-fixed axes, but precesses
and wobbles around the axis with the largest moment of inertia (MOI). The wobbling mode
arises from the rotation of a triaxial nucleus along the axis with the largest MOI
would quantum mechanically be disturbed by the rotations along the other two
axes due to the fact that the loss of axial symmetry for a triaxial nucleus leads to the
MOIs for three axes different from each other and thus makes the rotations
along the three axes all possible. Therefore, the wobbling motion is regarded as a
specific characteristic of a triaxial deformed nucleus, and together with the
chiral symmetry breaking~\cite{Frauendorf1997NPA, Frauendorf2001RMP, J.Meng2010JPG},
is considered as a unique fingerprint associated with the stable triaxial shape in nuclei.

The corresponding energy spectra for the wobbling mode are wobbling bands, a
sequence of $\Delta I=2$ rotational bands that build on different wobbling phonon
excitations~\cite{Bohr1975}. This kind of bands were first observed in odd-$A$ nucleus
$^{163}$Lu~\cite{Odegard2001PRL} and soon after were identified in triaxial strongly deformed
(TSD) region around $N=94$ in $^{161,163,165,167}$Lu~\cite{Jensen2002PRL, Jensen2002NPA,
Schonwasser2003PLB, Amro2003PLB, Hagemann2004EPJA, Bringel2005EPJA} and $^{167}$Ta~\cite{Hartley2009PRC}.
For even-even nuclei, the best example identified so far is $^{112}$Ru~\cite{Zhu2009IJMPE}.

On the theoretical side, the wobbling motion was studied by TRM~\cite{Bohr1975}
and particle rotor model (PRM)~\cite{Hamamoto2002PRC, Hamamoto2003PRC, Frauendorf2014PRC},
which are quantal models and exactly solved in the laboratory frame. Based on the framework
of cranking mean filed theory, the random phase approximation (RPA) method~\cite{Shimizu1995NPA,
Matsuzaki2002PRC, Matsuzaki2003EPJA, Matsuzaki2004PRC, Matsuzaki2004PRCa, Shimizu2005PRC,
Shimizu2008PRC, Shoji2009PTP}, the generator coordinate method after angular momentum projection
(GCM+AMP)~\cite{Oi2000PLB}, and the collective Hamiltonian~\cite{Q.B.Chen2014PRC} were
employed for investigating the wobbling motion. In addition, some approximate analytic
solutions, such as the harmonic approximation (HA) formula~\cite{Bohr1975, Frauendorf2014PRC}
and Holstein-Primakoff (HP) formula~\cite{Tanabe1971PLB, Tanabe2006PRC, Tanabe2008PRC},
were also obtained to describe the energy spectra and electromagnetic transition probabilities
of the wobbling motion.

Besides the energy spectra and electromagnetic transition probabilities, it would be very
interesting to study the angular momentum geometry pictures of the wobbling motions.
In Refs.~\cite{Hamamoto2002PRC, Jensen2002PRL, Jensen2002NPA}, the angular momentum components as
functions of spin for low lying wobbling bands are analyzed by PRM to study the nature of
the many-phonon wobbling excitations. Undoubtedly, such investigations for the angular momentum
geometry are essential to deeply understand the picture of wobbling motion. However, to our
knowledge, such investigations are still rather rare. Therefore, it is imperative to
systematically investigate the angular momentum geometry of wobbling motion.

In this paper, for simplicity, a simple triaxial rotor is taken as an example to
study the angular momentum geometry of wobbling motion by TRM. The energy spectra
and reduced quadrupole transition probabilities are obtained by exactly solving
the TRM Hamiltonian and compared with those approximate analytic results obtained by
HA~\cite{Bohr1975} and HP~\cite{Tanabe1971PLB} methods to evaluate the accuracy of the
approximations. The geometry of angular momentum as well as the $K$-distribution will
be analyzed in detail as functions of not only spin but also wobbling phonon number.

The paper is organized as follows. In Sec.~\ref{sec2}, the brief procedures of exactly and
approximately solving the TRM Hamiltonian are presented. The corresponding numerical details
used in the calculations are given in Sec.~\ref{sec3}. In Sec.~\ref{sec4}, the obtained energy
spectra and reduced quadrupole transition probabilities by exactly solving TRM are
shown and compared with those approximate results by HA and HP methods. The angular
momentum components and $K$-distribution evolution pictures as functions of spin and wobbling
phonon number are investigated. Finally, a brief summary is given in Sec.~\ref{sec5}.


\section{Theoretical framework}
\label{sec2}

\subsection{Triaxial rotor model}

  The triaxial rotor model was firstly introduced for the description of a rotating triaxial
  nucleus by Davydov and Filippov~\cite{Davydov1958NP} with the assumption that the nucleus
  has a well-defined potential minimum at a nonzero value of triaxial deformation parameter
  $\gamma$. Due to the anisotropy of a triaxial rotor, the nucleus can rotate around any of
  principal axis. The corresponding Hamiltonian thus reads as
  \begin{align}
  \label{eq4}
  \hat{H}=A_1\hat{I}_1^2+A_2\hat{I}_2^2+A_3\hat{I}_3^2,
  \end{align}
  in which $A_k$ are related to the MOIs of three principal axes $\mathcal{J}_k$ by
  $A_k=\displaystyle\frac{\hbar^2}{2\mathcal{J}_k}$, $k=1,2,3$. The details for exactly solving
  this Hamiltonian by diagonalization can be found in Ref.~\cite{Davydov1958NP} or in classical
  textbooks, e.g. Ref.~\cite{J.Y.Zeng2007book}. In this subsection, for completeness, a brief
  introduction is presented.

  The TRM Hamiltonian (\ref{eq4}) is invariant under $180^\circ$ rotation about the intrinsic
  principal axes ($D_2$ symmetry group). Thus the basis states $|IMK\alpha\rangle$ can be chosen
  as
  \begin{align}
  \label{eq17}
  &\quad|IMK\alpha\rangle\notag\\
  &=\frac{1}{\sqrt{2(1+\delta_{K0})}}\Big[|IMK\rangle+(-1)^I|IM-K\rangle\Big]\ \ \ (K\geq0),
  \end{align}
  where $|IMK\rangle$ is the Wigner $D$ function and $\alpha$ denotes other quantum numbers. The
  angular momentum projections onto the 3-axis in the intrinsic frame and $z$-axis in the laboratory
  are denoted by $K$ and $M$, respectively. Since the energy states are irrelevant to $M$, it
  is neglected in the following expressions.

  The TRM Hamiltonian (\ref{eq4}) can be written as
  \begin{align}
  \label{eq16}
  \hat H&=\Big[ \frac{1}{2}(A_1+A_2)(\hat I^2-\hat I_3^2)+A_3\hat I_3^2 \Big]\notag\\
        &+\frac{1}{4}(A_1-A_2)(\hat I_+^2+\hat I_-^2),
  \end{align}
  with the raising and lowering operators $\hat{I}_\pm=\hat{I}_1\pm i\hat{I}_2$. The first term of the Hamiltonian
  is diagonal while the second term only includes non-diagonal elements. The diagonal elements of
  $\hat{H}$ are
  \begin{equation}
  \langle IK|\hat H|IK\rangle=\frac{1}{2}(A_1+A_2)[I(I+1)-K^2]+A_3K^2.
  \end{equation}
  The second terms of Eq.~(\ref{eq16}) will cause the mixture of different states with $\Delta K=\pm2$
  and the corresponding matrix elements are
  \begin{align}
  \label{eq15}
  &\quad\langle IK|\hat H|IK\pm2\rangle\notag\\
  &=\frac{1}{4}(A_1-A_2)\sqrt{(I\mp K)(I\pm K+1)(I\mp K-1)(I\pm K+2)}
  \end{align}
  The energy eigenvalues and eigenstates for a given spin $I$ could be obtained by solving
  the eigen equation.

  The reduced electric quadrupole $(E2)$ transition probability is defined
  as~\cite{Bohr1975}
  \begin{equation}
  B(E2,I\rightarrow I^\prime)=\sum_{\mu KK^\prime}
  |\langle I^\prime K^\prime|\hat{\mathcal{M}}_{2\mu}^E|IK\rangle|^2
  \end{equation}
  using the quadrupole transition operator
  \begin{align}
  \hat{\mathcal{M}}_{2\mu}^{E}=\sqrt{\frac{5}{16\pi}}\hat Q_{2\mu},
  \end{align}
  of which the laboratory electric quadrupole tensor operator $\hat Q_{2\mu}$ are
  associated with the intrinsic ones via Wigner $D$ function
  \begin{align}
  \hat Q_{2\mu}=\mathcal{D}_{\mu0}^{2*}\hat{Q}'_{20}+(\mathcal{D}_{\mu2}^{2*}+\mathcal{D}_{\mu-2}^{2*})
  \hat{Q}'_{22}=\sum_{\nu}\mathcal{D}_{\mu\nu}^{2*}\hat{Q}'_{2\nu}.
  \end{align}
  The intrinsic quadrupole tensor operator are expressed by the intrinsic
  quadrupole $Q$ and triaxial deformation $\gamma$ as
  \begin{align}\label{eq18}
  Q_{20}^\prime=Q\cos\gamma, \quad Q_{22}^\prime=\frac{1}{\sqrt{2}}Q\sin\gamma.
  \end{align}

  With the solutions of the TRM, one can also investigate the angular momentum geometry. The
  expectation values of the squared angular momentum components for the total nucleus are calculated
  as $I_k=\sqrt{\langle \hat{I}_k^2 \rangle}$ ($k=1,2,3$). From the obtained angular momenta,
  the orientation of the total angular momentum, depicted by the polar angle $\theta$ and
  azimuth angle $\varphi$, can be extracted by
  \begin{align}
   I_1=J\sin\theta\cos\varphi, \quad I_2=J\sin\theta\sin\varphi, \quad I_3=J\cos\theta,
  \end{align}
  namely,
  \begin{align}\label{eq25}
   \theta=\cos^{-1}\Big(\frac{I_3}{J}\Big), \quad \varphi=\tan^{-1} \Big(\frac{I_2}{I_1}\Big),
  \end{align}
  where $J$ is the length of the total angular momentum $J=\sqrt{I(I+1)}$.

  \subsection{Approximate analytic expressions}
  \label{sec2subsec2}

  To find the analytic expressions for wobbling excitations, Bohr and Mottelson introduced
  the harmonic approximation for angular momentum operators~\cite{Bohr1975}. Taking
  3-axis the axis of largest MOI, in the HA, the third component of angular
  momentum $\hat{I}_3$ is approximated as a constant value $I$, and the angular momentum
  raising and lowering operators are expressed in terms of boson creation and annihilation
  operators
  \begin{align}\label{eq19}
    \hat{I}_+=\sqrt{2I}\hat{a}^{\dagger}, \quad \hat{I}_-=\sqrt{2I}\hat{a},
  \end{align}
  with $[\hat{a},\hat{a}^{\dagger}]\approx 1$. With this assumption, the energy spectra
  can be expressed as the sum of two parts~\cite{Bohr1975}
  \begin{equation}\label{eq22}
    E(n,I)=A_3I(I+1)+(n+\frac{1}{2})\hbar\omega.
  \end{equation}
  The first one is the rotation energy with respect to the rotational axis with spin $I$ and
  the second one is a harmonic oscillation energy, i.e., wobbling energy, with wobbling phonon
  number $n$. In details, the wobbling phonon number $n$ characterizes the wobbling motion of
  the axes with respect to the direction of $I$ and the wobbling frequency $\hbar\omega$ is
  determined by the MOI as~\cite{Bohr1975}
  \begin{equation}\label{eq1}
  \hbar\omega=\sqrt{\alpha^2-\beta^2}I=2I\sqrt{(A_2-A_3)(A_1-A_3)},
  \end{equation}
  with
  \begin{align}\label{eq20}
  \alpha=A_2+A_1-2A_3,\quad \quad \beta=A_2-A_1.
  \end{align}

  Since the wobbling motion is of harmonic vibration character, one can extract the
  ``kinetic'' and ``potential'' terms by combination of the wobbling boson creation
  and annihilation operators. They are respectively written as $(A_2-A_3)I_2^2$
  and $(A_1-A_3)I_1^2$ when the nucleus rotates around the 3-axis.

  With the HA, the intraband $B(E2)$ transition probabilities are approximated
  as~\cite{Bohr1975}
  \begin{equation}\label{eq24}
  B(E2;n,I\rightarrow n,I-2)\approx\frac{5}{16\pi}Q_{22}^{\prime2},
  \end{equation}
  and the interband ones~\cite{Bohr1975}
  \begin{align}\label{eq2}
  &B(E2;n,I\rightarrow n+1,I-1)\notag\\
  &\quad=\frac{5}{16\pi}\frac{n+1}{I}\Big(\sqrt{3}Q_{20}^\prime y
  +\sqrt2Q_{22}^\prime x\Big)^2,
  \end{align}
  and
  \begin{align}\label{eq3}
  &B(E2;n,I\rightarrow n-1,I-1)\notag\\
  &\quad=\frac{5}{16\pi}\frac{n}{I}\Big(\sqrt{3}Q_{20}^\prime x
  +\sqrt2Q_{22}^\prime y\Big)^2,
  \end{align}
  where the coefficients are defined as
  \begin{align}
  \left.
  \begin{aligned}
  x & \\
  y & \\
  \end{aligned}
  \right\}
  &=\sqrt{\frac{1}{2}\Big(\frac{\alpha}{\sqrt{\alpha^2-\beta^2}}\pm1\Big)}.
  \end{align}
  The definitions for the intrinsic quadrupole moments $Q_{20}^\prime$ and $Q_{22}^\prime$
  are the same as Eq.~(\ref{eq18}).

  In the HA method, the angular momenta are expanded by boson operators only in zeroth
  order. Tanabe and Sugawara-Tanabe further applied the Holstein-Primakoff (HP)
  transformation~\cite{Holstein1940PR} to the TRM Hamiltonian~\cite{Tanabe1971PLB,
  Tanabe2006PRC}. The HP transformation expresses the angular momentum operators
  in terms of boson creation and annihilation operators as
  \begin{align}
  \hat{I}_3 &= I-\hat{n}, \notag \\
  \hat{I}_+ &= \hat{a}^\dagger \sqrt{2I} \sqrt{1-\frac{\hat{n}}{2I}},\notag\\
  \hat{I}_- &= \sqrt{2I} \sqrt{1-\frac{\hat{n}}{2I}}\hat{a}.
  \end{align}
  with $\hat{n}=\hat{a}^\dagger\hat{a}$. When $I\gg 1$ and $\hat{n}$ is small, the square
  roots $\displaystyle\sqrt{1-\frac{\hat{n}}{2I}}$ can be expanded as Taylor series with
  respect to $\hat{n}=0$. If the expansion is in zeroth order, the HP transformation
  degenerates to HA method~(\ref{eq19}). By taking into account the effect of the next-to-leading
  order term, the wobbling energy spectra are finally given by~\cite{Tanabe1971PLB, Tanabe2006PRC}
  \begin{align}\label{eq21}
  E(n,I)&=A_3I(I+1)\notag\\
  &+\Big(I\sqrt{\alpha^2-\beta^2}+\frac{1}{2}\sqrt{\alpha^2-\beta^2}-\frac{\alpha}{2}\Big)
  (n+\frac{1}{2})-\frac{\alpha}{2}n^2.
  \end{align}
  The definitions of $\alpha$ and $\beta$ can be found in Eq.~(\ref{eq20}).

\section{Numerical details}
\label{sec3}

  In the following, the wobbling motion for a triaxial rotor with $A_1=2A_2=6A_3=0.06~$MeV$/\hbar^2$
  is discussed. The MOI values of three axes are thus $\mathcal{J}_1=8.33~\hbar^2/\rm MeV$,
  $\mathcal{J}_2=16.66~\hbar^2/\rm MeV$, and $\mathcal{J}_3=50.00~\hbar^2/\rm MeV$.
  Namely, the largest and the smallest MOI axes are respective 3-axis and 1-axis. The
  assumption for the MOI is the same as Ref.~\cite{Frauendorf2014PRC}. For the $B(E2)$
  calculations, the intrinsic quadrupole moment $Q_0$ is taken as $\sqrt{16\pi}e\rm b$
  and the triaxial deformation parameter is assumed to be $\gamma=30^\circ$.


\section{Results and discussion}
\label{sec4}

In the following, we first present the energy spectra and electromagnetic
transition probabilities obtained by TRM and compare them with those by
HA and HP method. Then, the angular momenta evolution picture will be
discussed in details. Finally, the microscopic understanding for this
evolution will be shown.

\subsection{Energy spectra}
\label{aa}

\begin{figure}[!ht]
  \begin{center}
    \includegraphics[width=8 cm]{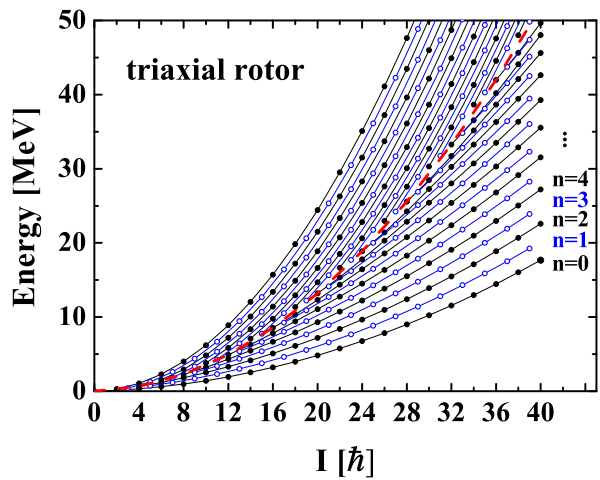}
    \caption{(Color online) Energy spectra as functions of spin $I$ for a triaxial rotor with
    $A_1=2A_2=6A_3=0.06~$MeV$/\hbar^2$ calculated by TRM. The full dots belong to the states
    with signature $\alpha=0$, while empty dots to signature $\alpha=1$. The approximate quantum number
    $n$ ($n=0,1,2,\cdots$) is introduced to label the energy states in accordance with the
    increasing order of their energies. Similar figure can be found in Ref.~\cite{Frauendorf2014PRC}.}\label{fig1}
  \end{center}
\end{figure}

In Fig.~\ref{fig1}, the energy spectra as functions of spin $I$ calculated by the TRM
are shown. The sates with signature $\alpha=0$ and $\alpha=1$ are respectively plotted
by full and empty dots. The approximate quantum number $n$ ($n=0,1,2,\cdots$) is introduced
to label the energy states in order of their energies, and the energy for each state is
denoted by $E(n,I)$. Since a triaxial nucleus possesses $D_2$ symmetry,
the energy spectra are restricted to the states with $(-1)^n=(-1)^I$~\cite{Bohr1975}, i.e.,
only even spins appear for even-$n$ states while only odd spins appear for odd-$n$ states.
Thus, there are $I/2+1$ states for a given even spin $I$, while $(I-1)/2$ states for
odd spin $I$.

The obtained energy spectra can be divided into two different groups,
which are distinguished by the manners of linking. One of them consists of the lines linking
states with the same $n$ for $n$-small bands. With increasing spin, the energy spacing between
neighboring lines become more and more equivalent. This is the classical wobbling regime,
which corresponds to the wobbling motion around the axis with the largest MOI
(i.e., 3-axis). The other one consists of the lines linking states with the same $I-n$
for $n$-large bands, which corresponds to the wobbling motion around the axis with the
smallest MOI (i.e., 1-axis). Between them, it is the separatrix region. It has
the energy of the unstable uniform rotation about the axis with the intermediate
MOI.

\begin{figure}[!ht]
  \begin{center}
    \includegraphics[width=8 cm]{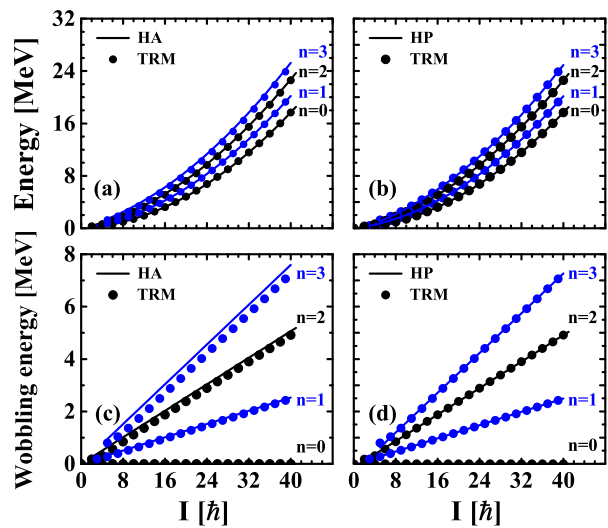}
    \caption{(Color online) The energy spectra and wobbling energies for the four lowest wobbling
  bands calculated by TRM are compared with those by HA (a, c) and HP (b, d) formulas.}\label{fig2}
  \end{center}
  \end{figure}

  To examine the accuracy of HA (\ref{eq22}) and HP (\ref{eq21}) formulas, the energy spectra
  and wobbling energies as functions of spin for the four lowest wobbling bands obtained by
  them are in comparison with the exact solutions by TRM and shown in Fig.~\ref{fig2}. The
  wobbling energies, defined as the energy differences between the excited states and yrast
  state, are extracted from the obtained energy spectra and for a given spin $I$, are
  calculated as
  \begin{align}
  E_{\textrm{wob}}=E(n,I)-E(0,I)
  \end{align}
  for even $I$, and
  \begin{align}
  E_{\textrm{wob}}=E(n,I)-\frac{1}{2}[E(0,I-1)+E(0,I+1)]
  \end{align}
  for odd $I$.

  It can be seen that the wobbling energies obtained by the three methods all
  increase with the spin. This phenomenon can be microscopically understood
  from the fact that the stiffness of the collective potential increases with
  spin~\cite{Q.B.Chen2014PRC}. Both the HA and HP methods can reproduce the TRM results
  for both energy spectra and wobbling energies over the whole spin range for small $n$
  wobbling bands. However, with the increasing of $n$, the HA gradually deviates
  from the TRM results. For example, the HA results are larger than the TRM results
  about 150 keV for $n=2$ wobbling band and about 350 keV for $n=3$ wobbling
  band. This indicates that the wobbling motion is no longer with good harmonic vibrator
  character if $n$ becomes too large and the anharmonicity of wobbling motions begins to play
  an important role~\cite{Jensen2002PRL, Oi2005JPG, Oi2006PLB}. The occurrence of anharmonicity
  can be described by taking into account the higher order terms in the expansion of angular
  momenta. This statement is supported by the fact that HP formula reproduces the exact solutions
  perfectly. The energy differences between HP and TRM results are less than 30 keV even for
  $n=3$ wobbling band.

  To investigate the effects of the high order terms quantitatively, the energy differences
  obtained by the HA and HP methods are calculated
  \begin{align}
  &E_{\textrm{HA}}(n,I)-E_{\textrm{HP}}(n,I)\notag\\
  &\quad =\frac{\alpha}{2}n^2-\Big(\frac{1}{2}\sqrt{\alpha^2-\beta^2}
  -\frac{\alpha}{2}\Big)(n+\frac{1}{2}).
  \end{align}
  It depends on wobbling phonon number $n$ while is independent of spin, i.e., the roles played
  by the next-to-leading term are different for different $n$ while are the same for different
  spin. It is obvious that with the increasing of $n$, the energy difference between HP and HA
  becomes larger, which indicates that the high order terms become more important and derive
  the wobbling motion deviates from harmonic vibrator behavior.

\subsection{Quadrupole transition probability}
\label{bb}

\begin{figure}[!ht]
  \begin{center}
  \includegraphics[width=8 cm]{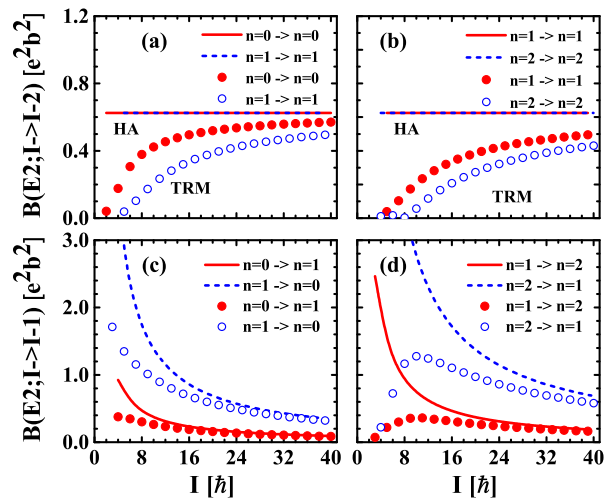}
  \caption{(Color online) The intraband and interband $B(E2)$ values for $n=0$, $1$ (a, c)
  and $n=1$, $2$ (b, d) bands calculated by TRM in comparison with those by HA
  formula~\cite{Bohr1975}.}\label{fig3}
  \end{center}
  \end{figure}

  In Fig.~\ref{fig3}, the intraband and interband $B(E2)$ values for $n=0$, $n=1$,
  and $n=2$ bands calculated by TRM in comparison with those by HA
  formula~\cite{Bohr1975} are shown. For the intraband $B(E2)$ values, the HA results
  given by Eq.~(\ref{eq24}) are constants, which are independent of spin $I$ and wobbling
  phonon number $n$. However, as shown in Fig.~\ref{fig3}, the TRM results depend on not
  only spin but also wobbling phonon number. For each wobbling band $n$, the intraband $B(E2)$
  values gradually increase with increasing spin and finally come close to the HA results at very
  high spin. For a given spin, with the increasing of $n$, the corresponding $B(E2)$ values
  decrease. Therefore, one can conclude that the HA formula is a good approximation only
  at very high spin and small $n$. This is expected since the approximation for vector addition
  coefficients in $E2$ matrix elements, which is valid only at very high spin, has been adopted
  when deriving the HA formula~\cite{Bohr1975}.

  For interband $B(E2)$ values, the HA results exhibit decreasing trend with respect to
  spin for given $n$ and increasing trend with respect to $n$ for given spin, which are
  determined by the factor $n/I$ in the HA formulas (\ref{eq2}) and (\ref{eq3}). It can be further
  seen that the $B(E2;n,I\rightarrow n+1,I-1)$ values are smaller than the $B(E2;n+1,I\rightarrow
  n,I-1)$ values over the whole spin region since the bracketed terms in the Eq.~(\ref{eq2})
  are smaller than the bracketed terms in the Eq.~(\ref{eq3}). For TRM results, the decreasing
  trends with respect to spin can also be found at high spin region. While at low spin region,
  contrary to the HA results, the increasing trend is shown for the interband transition
  between $n=1$ and $n=2$ bands. For example, the $B(E2;2,I\rightarrow 1,I-1)$ values increase
  rapidly with spin at $I\leq 10\hbar$.

  It can be also seen that the HA results in general overestimate the TRM results.
  In particular, this overestimation becomes larger with the increasing of
  wobbling phonon number $n$. For a given $n$, the TRM results gradually come close to the HA
  results as spin increases. Therefore, similar to the case of intraband transition
  probabilities, the interband transition probabilities obtained by HA formula are
  good approximations for those by TRM only at high spin region and small $n$.

  Comparing the interband transition with the intrband transition, it is seen that
  the strength of interband $B(E2;n,I\rightarrow n\pm1,I-1)$ is smaller than that for the
  intraband $B(E2;n,I\rightarrow n,I-2)$ without changing the wobbling phonon number $n$ at
  high spin region, by a factor of order $\displaystyle\frac{n}{I}$, which represents the
  square of the amplitude of the wobbling motion~\cite{Bohr1975}. This is the expected
  features of the wobbling bands, which corresponds to a sequence of one-dimensional
  rotational trajectories with strong $E2$ transition along the trajectories and much
  weaker transitions connecting the different trajectories~\cite{Bohr1975}.

\subsection{Angular momentum}
\label{c}

\begin{figure}[!ht]
  \begin{center}
  \includegraphics[width=7 cm]{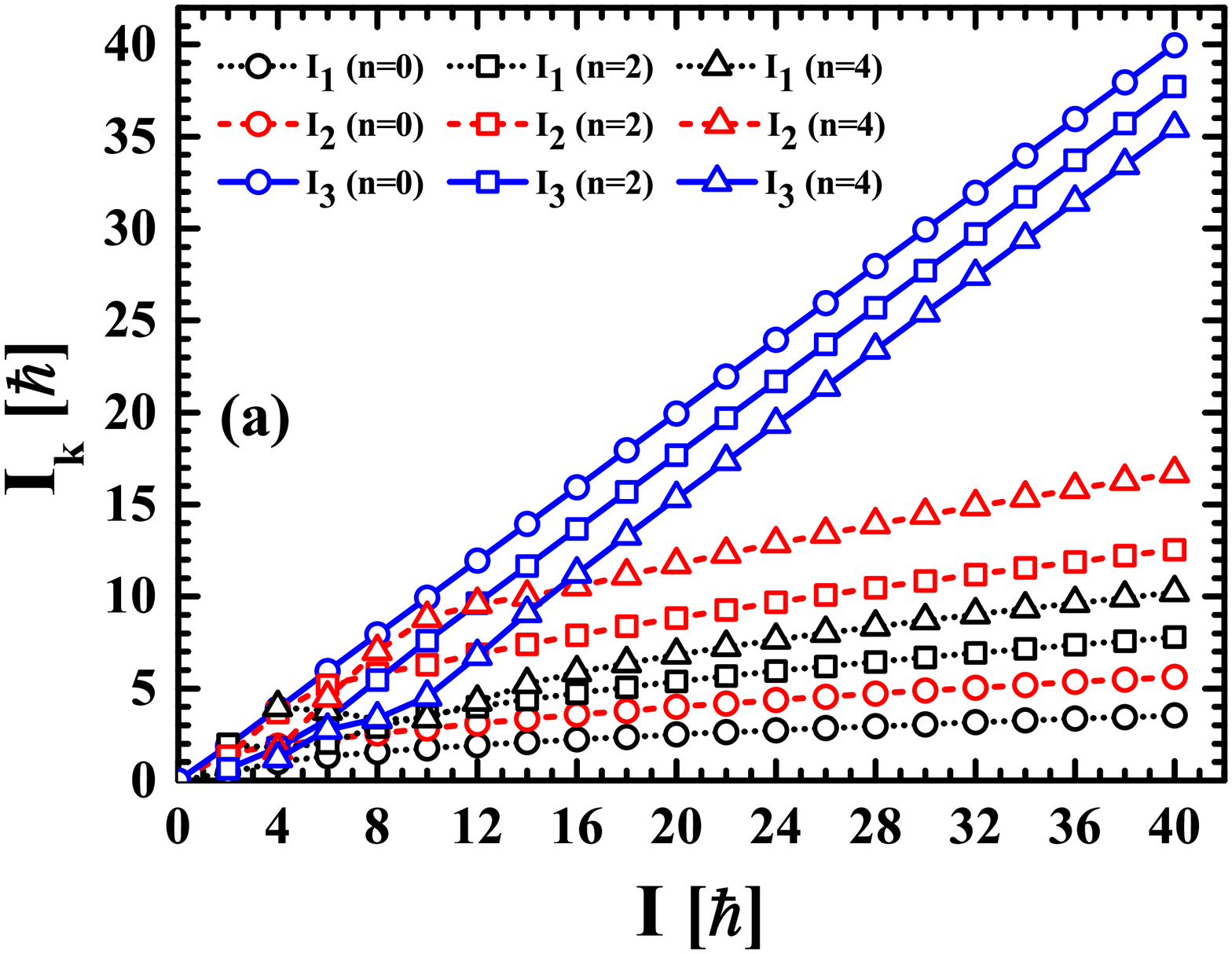}\quad
  \includegraphics[width=7 cm]{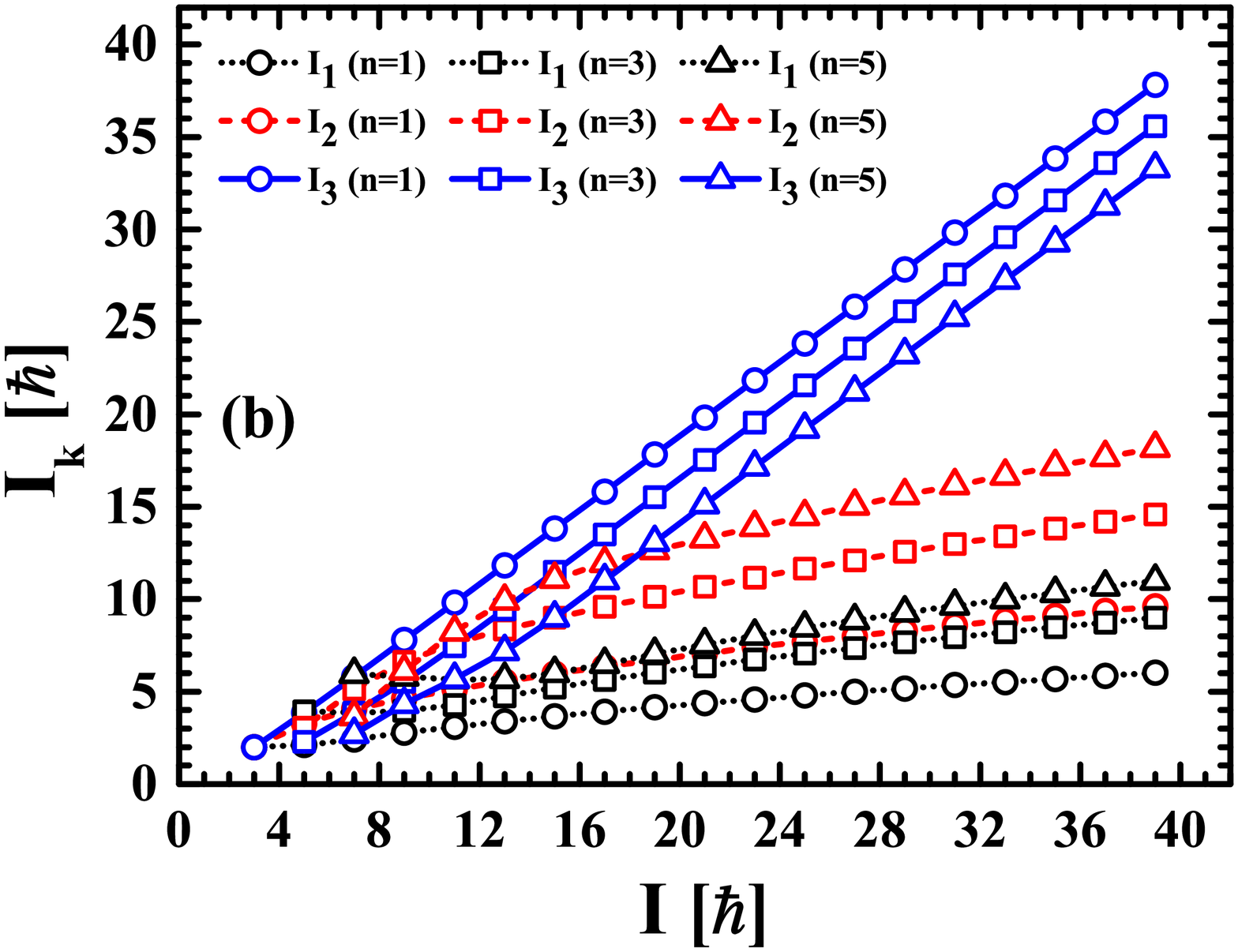}
  \caption{(Color online) The root mean square components along the 1, 2, and 3 axis of the total
   nucleus $I_k=\sqrt{\langle \hat{I}_k^2\rangle}$ calculated as functions of spin $I$ by TRM for
   (a) even $n$ wobbling bands ($n=0$, $2$, $4$) and (b) odd $n$ wobbling bands ($n=1$, $3$,
   $5$).}\label{fig4}
  \end{center}
  \end{figure}

  To understand the wobbling geometry picture, it is necessary to investigate the evolutions
  picture of angular momentum components with respect to spin $I$ and wobbling phonon number
  $n$. In Fig.~\ref{fig4}, the angular momentum components $I_k$ as functions of spin for $n=0$-$5$
  bands calculated by TRM are shown. For $n=0$ band, all of the three components of the angular
  momentum $I_k$ of the triaxial rotor gradually increase with spin increases. However, the roles
  played by them are quite different. The angular momentum favors aligning along the 3-axis with the
  largest MOI to obtain the lowest energy. The value of $I_3$ is nearly equal to $I$.
  The contributions from the other two axes to the total angular momentum are thus very small due
  to the conservation of the angular momentum
  \begin{align}
  I_1^2+I_2^2+I_3^2=J^2=I(I+1).
  \end{align}
  It is worth pointing out that the angular momentum along the 3-axis and the other two axes
  (1-, 2-axis) respectively contribute the rotational energy and the zero energy of the wobbling
  motion for the wobbling energy spectra.

  With the increasing of the wobbling phonon number $n$, the angular momentum shows a tendency
  to gradually deviate from the 3-axis. The value of $I_3$ is nearly equal to $I-n$ at large
  spin region. Meanwhile, the values of the 1- and 2- components of the angular momentum become
  larger and larger, which indicates that the wobbling motion plays a more and more
  important role.

  It should be noted that when the wobbling phonon number $n$ is larger than 2, the three
  components of the angular momentum will present a strong competitive behavior.
  There exhibits a clear picture that with the increasing of spin the maximal component of the
  angular momentum changes from $I_1$ to $I_2$ and finally to $I_3$. Here, the evolution
  of the angular momentum with spin for $n=4$ band will be taken as an example to
  illustrate this picture. At spin $I=4\hbar$, the $n=4$ state is the largest energy
  state as shown in Fig.~\ref{fig1} and the angular momentum mainly aligns along the 1-axis,
  which corresponds to the wobbling motion along the axis with the smallest MOI.
  At spin region $6\leq I\leq 14\hbar$, the angular momentum turns to mainly align along
  the 2-axis with intermediate MOI. It can be seen from Fig.~\ref{fig1} that these states
  lie around the separatrix region. When spin $I>16\hbar$, the angular momentum finally mainly
  aligns along the 3-axis, which corresponds to the wobbling motion along the axis with the
  largest MOI. As shown in Fig.~\ref{fig1}, these states are beneath the separatrix line
  and become further away from it as spin increases.

\begin{figure}[!ht]
  \begin{center}
  \includegraphics[width=8 cm]{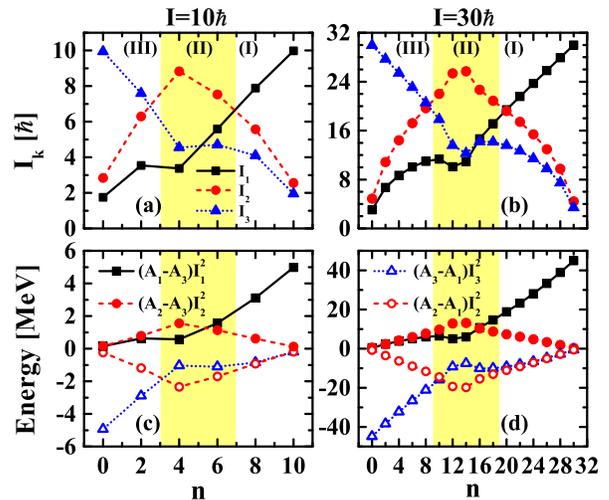}
  \caption{(Color online) The root mean square components along the 1, 2, and 3 axis of the total
   angular momentum $I_k=\sqrt{\langle \hat{I}_k^2\rangle}$ calculated by TRM and their corresponding
   contributions to the ``kinetic'' and ``potential'' terms of wobbling energy as functions of
   wobbling phonon number $n$ for $I=10\hbar$ (a, c) and $I=30\hbar$ (b, d).}\label{fig5}
  \end{center}
  \end{figure}

  It has been emphasized that there exists complex competitions among three components
  of angular momenta with spin $I$ in the same $n$ wobbling bands. It is also
  interesting to investigate the evolution picture with respect to the wobbling
  phonon number for a given spin $I$. Therefore, in Fig.~\ref{fig5}, the root mean square
  components along the 1, 2, and 3 axis of the total angular momentum $I_k$ calculated by TRM
  as functions of wobbling phonon number $n$ are shown respectively for $I=10\hbar$
  (Fig.~\ref{fig5}(a)) and $I=30\hbar$ (Fig.~\ref{fig5}(b)).

  For $I=10\hbar$, in general, the $I_3$ decreases while the $I_1$ increases as wobbling phonon
  number $n$ increases, which indicates that the angular momentum generally has a tendency to
  move away from 3-axis at small $n$ and to 1-axis at large $n$. However, $I_3$ shows a
  slightly increasing behavior when $n$ increases from 4 to 6. The angular momentum of $n=6$
  state is closer to 3-axis than that of for $n=4$ state although the difference between them
  is small. This indicates that the angular momentum does not always steadily deviate from the
  3-axis with increasing of $n$. It is further noted that $I_1$ also exhibits an abnormal behavior. When $n$
  increases from 2 to 4, it decreases about $0.2\hbar$. Thereafter, it increases linearly with
  the increasing of $n$. Differing from the behaviors of $I_1$ and $I_3$, $I_2$
  increases first at $n\leq 4$ and then decreases. Therefore, the strong competition
  among the three components of angular momentum is clearly presented. When $n=0$ and $2$, the $I_3$
  is largest and $I_2>I_1$. This region is labeled as (III) in the figure. While when $n=4$ and $6$,
  $I_2$ becomes the largest component and $I_1$ and $I_3$ compete with each other. This region is
  labeled as (II). Finally, when $n$ increases to $8$ and $10$, $I_1$ begins to play the main role
  while $I_3$ turns to be the smallest. Correspondingly, this region is labeled as (I).
  For $I=30\hbar$, the evolution behaviors of three components of angular momentum are very similar to
  those for $I=10\hbar$. The abnormal behaviors exhibited by $I_3$ and $I_1$ happen at $n=14, 16$ and
  $n=10, 12$. For $I_2$, it reaches to maximal value at $n=14$. With increasing of $n$, the three
  components of angular momentum present a strong competitive behavior as well. Similar to the case of
  $I=10\hbar$, (III), (II), and (I) are labeled according to the maximal components of the angular momentum.

  It can be thus concluded that with the increasing of the wobbling phonon number $n$ for a certain
  spin $I$, the simple triaxial rotor undergoes three different kinds of regions, as labeled
  (III), (II), and (I) regions in Fig.~\ref{fig5}. In the region (III), $I_3$ is the largest,
  corresponding to the wobbling motion around the 3-axis with the largest MOI.
  This is further supported by the nearly equivalent contributions of ``kinetic'' term $(A_2-A_3)I_2^2$
  and ``potential'' term $(A_1-A_3)I_1^2$ in this region, as shown in the Figs.~\ref{fig5}(c) and (d).
  In addition, in this region, $I_3$ is approximately linear to $n$ with
  $I_3\approx I-n$. In the region (I), however, $I_1$ is the largest, which corresponds to the
  wobbling motion around the 1-axis with the smallest MOI. Similarly, this is also
  supported by the equivalent contributions of ``kinetic'' term $(A_3-A_1)I_3^2$ and ``potential''
  term $(A_2-A_1)I_2^2$ in this region, as shown in the Figs.~\ref{fig5}(c) and (d).
  Similar to $I_3$, $I_1$ is also approximately linear to $n$ with $I_1\approx n$ for even spin
  and $I_1\approx n+1$ for odd spin. In the transitional region (II), which corresponds to the
  separatrix region in Fig.~\ref{fig1}, the $I_1$, $I_2$ and $I_3$ exhibit a completely
  different behavior from those shown in other two regions and fierce competitions
  among the three components of angular momentum is clearly shown. The fierce competitions
  might be associated with the phenomenon, as shown in Fig.~\ref{fig1}, that the energy in this
  region increases slowly compared with those in other two regions as $n$ increases. In addition,
  although the $I_2$ is largest, there is not wobbling motion around 2-axis with the intermediate
  MOI since the ``kinetic'' term $(A_1-A_2)I_1^2$ is positive ($A_1>A_2$) while the ``potential''
  term $(A_3-A_2)I_3^2$ negative ($A_3<A_2$). This conclusion is consistent with that obtained
  by tops-on-top model~\cite{Tanabe2014PTEP}.

\begin{figure}[!ht]
  \begin{center}
  \includegraphics[width=7 cm]{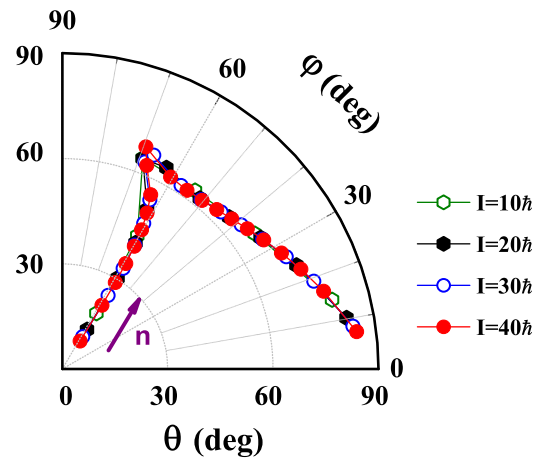}
  \caption{(Color online) The calculated orientation of the total angular momentum, depicted by
  the polar angle $\theta$ and azimuth angle $\varphi$, as a function of wobbling phonon number $n$
  for $I=10$, $20$, $30$ and $40\hbar$. The $\theta$ and $\varphi$ are shown in a polar coordinate system,
  in which $\theta$ is the radial coordinate while $\varphi$ the angular coordinate. The arrow
  indicates the increasing direction of wobbling phonon number.}\label{fig6}
  \end{center}
  \end{figure}

  In order to further clear draw the angular momentum evolution pictures, in Fig.~\ref{fig6}, the
  orientation of the total angular momentum, depicted by the polar angle $\theta$~(\ref{eq25}) and
  azimuth angle $\varphi$~(\ref{eq25}), as a function of wobbling phonon number $n$ for $I=10$, $20$,
  $30$ and $40\hbar$ are illustrated.

  It is clearly seen that the series of points with the wobbling phonon number $n$ for all
  spin $I$ lies on the same line, which indicates that no matter what spin $I$ is, the orientations
  of angular momentum vectors are along the same evolution track with the increasing of $n$.
  When $n$ is small, the $\varphi$ is almost invariable while the $\theta$ increases as $n$ increases,
  which suggests that the angular momentum vector lies almost in the same plane $\varphi\approx 60^\circ$
  but gradually deviates from the 3-axis as $n$ increases. The reason of angular momentum
  being in the plane $\varphi\approx 60^\circ$ can be understood according to the
  ``kinetic'' and ``potential'' terms in the TRM Hamiltonian as shown in the
  Figs.~\ref{fig5}(c) and (d). Since the ``kinetic'' term $(A_2-A_3)I_2^2$ and ``potential'' term $(A_1-A_3)I_1^2$
  are almost equivalent for small $n$, it indicates that $\varphi=\tan^{-1}\Big(\displaystyle\frac{I_2}{I_1}\Big)$
  approximates as $\tan^{-1}\sqrt{\displaystyle\frac{A_2-A_3}{A_1-A_3}}$. Thus $\varphi$ is independent on
  $\theta$ but determined by the three MOIs at small $n$. According
  to this formula, one can obtain $\varphi\approx 60^\circ$ in the present discussed system. It
  is noted that these energy states correspond to the wobbling motion along the 3-axis and mainly
  lie in the region (III) as shown in Fig.~\ref{fig5}. As $n$ increases, both $\theta$ and $\varphi$
  increase, which suggests that the angular momentum not only continues deviating from the 3-axis,
  but also starts moving to the 2-axis. However, the increasing trend of $\varphi$ will not be
  incessant but will interrupt up to a critical point. This critical point corresponds to the maximal
  values of $I_2$, as shown in the Fig.~\ref{fig5}(a) and (b). After that, the $\varphi$, together
  with $\theta$, begins to decrease. It indicates that the angular momentum moves to the 1-axis,
  and also moves slightly to the 3-axis in the meantime. The points around the critical point
  correspond to the transition region (II) shown in Fig.~\ref{fig5}. After crossing this regions,
  $\varphi$ decreases while $\theta$ increases with the increasing of $n$, moving
  to region (I) shown in Fig.~\ref{fig5}. Similar to the cases of small $n$, according to
  Figs.~\ref{fig5}(c) and (d), in this region it has $\displaystyle\frac{I_3}{I_2}
  =\sqrt{\frac{A_3-A_1}{A_2-A_1}}$, which implies that the angular momentum vectors for
  large $n$ lie also in the same plane that is perpendicular to the 2-3 plane.

  According to the above discussions, the evolution track of angular momenta with respect
  to the wobbling phonon number presented in Fig.~\ref{fig6} is determined only by the
  three MOIs of the triaxial rotor. As already pointed out above, the
  evolution track is unique and irrespective of what spin is. From this point of view,
  this evolution track may be regarded as a characteristic feature for a rotating triaxial rotor.
  Just along this evolution track, the triaxial rotor changes its rotation along the axis with
  the largest MOI to the axis with the smallest MOI.

\subsection{$K$-distribution}

  In the above, the angular momentum geometry of wobbling motion for the triaxial rotor
  are investigated via the expectation values of the three components of the angular momentum.
  It is interesting to further investigate the angular momentum geometry pictures from a
  microscopic perspective by the obtained wave functions. In the TRM, the wave functions can be
  directly reflected by the so called $K$-distributions, i.e., the probability distributions of
  projection of total angular momentum on the 1-, 2- and 3- axis. In the literatures, by the PRM,
  such $K$-distributions have been extensively applied to study the angular momentum geometry of
  the nuclear chiral modes in triaxial nuclei~\cite{B.Qi2009PLB, B.Qi2009PRC, Q.B.Chen2010PRC,
  B.Qi2011PRC}.

\begin{figure}[!ht]
  \begin{center}
  \includegraphics[width=7.5 cm]{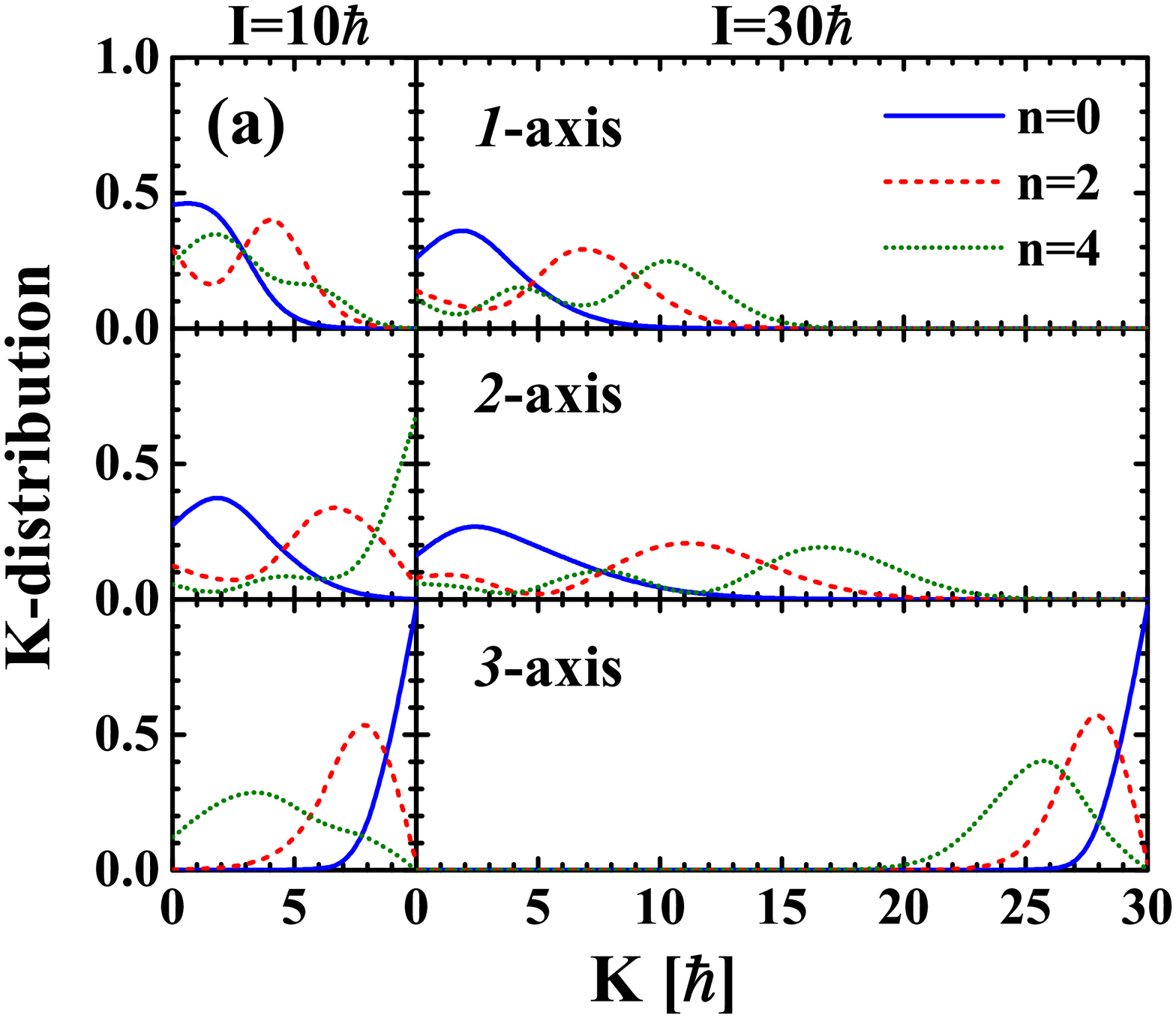}
  \includegraphics[width=7.5 cm]{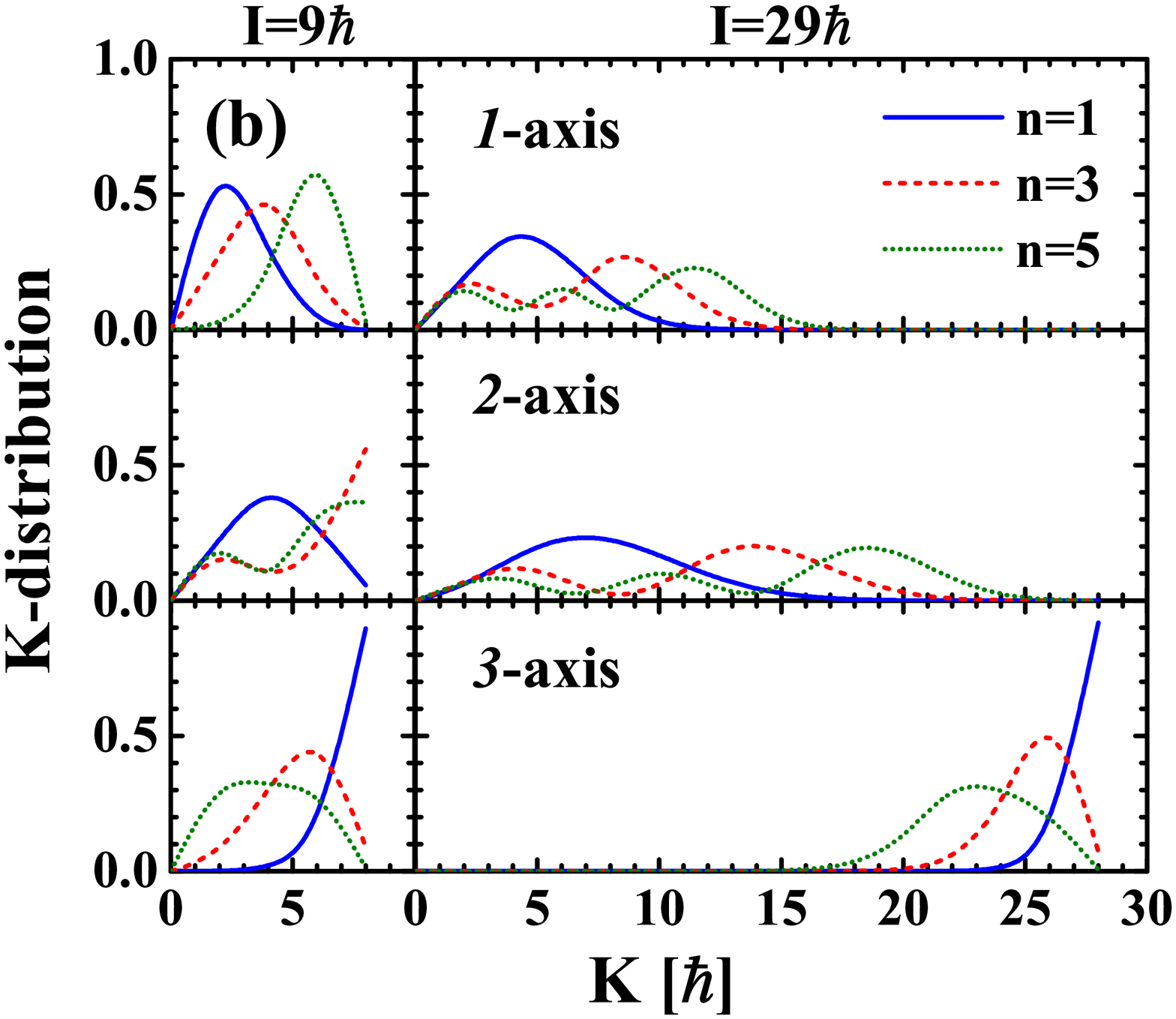}
  \caption{(Color online) The probability distributions for projection of total angular momentum on the 1-,
  2- and 3- axis in TRM for (a) $n=0$, $2$, $4$ states with $I=10$ and $30\hbar$ and (b)
  $n=1$, $3$, $5$ states with $I=9$ and $29\hbar$.}\label{fig7}
  \end{center}
  \end{figure}

  In the following, the $K$-distributions for the $n=0$, $2$, $4$ states with $I=10$ and $30\hbar$
  and $n=1$, $3$, $5$ states with $I=9$ and $29\hbar$, shown in Fig.~\ref{fig7}, are taken as
  examples to microscopically understand the angular momenta geometry.

  It can be seen that for $n=0$ states, the maximal probability for the 3-axis appears at $K_3=I$
  for both $I=10$ and $30\hbar$, while the probabilities for the other $K_3$ values almost
  vanish. This is expected since the angular momentum mainly along, as shown in Fig.~\ref{fig5}, 3-axis.
  The probability distributions with respect to the 2- and 1-axis have peaks near $K_2=2\hbar$ and
  $K_1=2\hbar$ for both $I=10$ and $30\hbar$. The reason why these peaks do not appear at $K_2=0$
  and $K_1=0$ is attributed to the nature of three dimensional rotation of the triaxial rotor, where
  three angular momentum components $I_k$ all have contributions to the total spin. As the MOI
  of 2-axis is larger than that of 1-axis, the angular momentum is closer to 2-3 plane than 1-3 plane.
  Thus the tunneling of angular momentum with respect to the 2-3 plane is easier than to 1-3 plane.
  This phenomenon can be reflected by the probability for $K_1=0$ is larger than that for $K_2=0$.

  As $n$ increases, for large spin, e.g., $I=29$ and $30\hbar$, the peak of the probability on 3-axis
  moves away from the $K=I$ to $K=I-n$ corresponding to a deviation from the 3-axis, which suggests
  that the distributions from $I_1$ and $I_2$ become more important. Meanwhile, for $K_1$ and $K_2$,
  more peaks appear and it is expected that the number of peaks is $n+1$ if the $K$
  distributions are plotted also for negative $K$ values. It is noted that for the even $n$ and odd $n$
  states, the probabilities on the 1- and 2- axis are respective nonzero and zero
  at $K=0$, which indicates that their corresponding wave functions are symmetric and antisymmetric in
  the projections on the 1- and 2- axis. We should bear in mind that all the presented states for large spin
  lie in the region (III), where the wobbling motions happen along the axis with largest MOI.
  Once the states lying in the other two regions are involved, the $K$ distributions would exhibit a
  different behavior. This can be seen for small spin when $n$ is not too large, e.g., $n=4$ state
  with $I=10\hbar$, which lies in the region (II). Therefore, further investigation for the evolutions
  of $K$ distribution with respect to the wobbling phonon number $n$ is necessary.

  \begin{figure}[!ht]
  \begin{center}
  \includegraphics[width=6 cm]{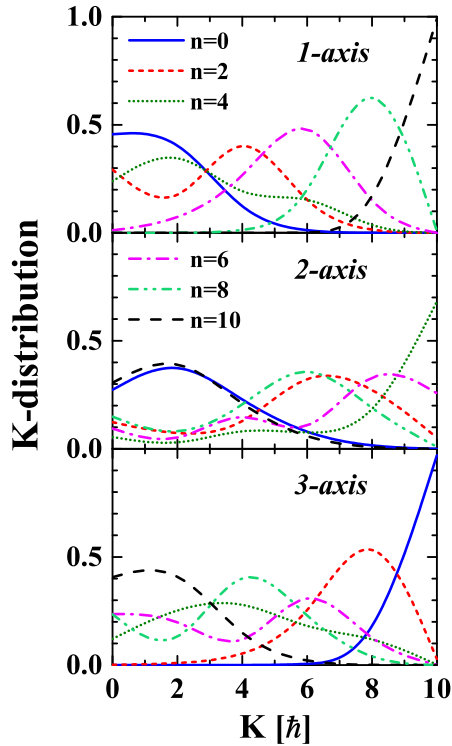}
  \caption{(Color online) Same as Fig.~\ref{fig7}, but for the $n=0$-$10$ states with $I=10\hbar$.}\label{fig8}
  \end{center}
  \end{figure}

  In Fig.~\ref{fig8}, the $K$ distributions for the $n=0$-$10$ states are shown.
  For $n=4$, the probability for $K_2=I=10\hbar$ is much larger than that for
  other $K_2$. Comparing with the region (II) in Fig.~\ref{fig5}, this corresponds
  to the situation that $I_2$ is the largest angular momentum components and the
  angular momentum mainly aligns along the 2-axis. In this case, the probability
  distributions with respect to the 1- and 3-axis have peaks near $K_1=2\hbar$
  and $K_3=4\hbar$. As $n$ increases, the peak of the probability on 1-axis gradually
  moves towards to $K_1=I$ and the peaks for each $n$ state appear nearly at $K_1=n$,
  which indicates a tendency of the angular momentum moving to the 1-axis.
  This evolution picture is consistent with the evolution picture of angular
  momentum shown in Fig.~\ref{fig5}.

  It is interesting to note that the $K$ distribution on the 1-axis and 3-axis for
  the $n=8$ and $n=10$ states are respectively similar to those distributions on the
  3-axis and 1-axis for the $n=2$ and $n=0$ states. Moreover, the $K$ distribution on
  the 2-axis for $n=0$ and $n=10$ states as well as $n=2$ and $n=8$ states are also
  very similar. These similar behaviors are justified since the wobbling motions happen
  both along 1- and 3- axis. These similarity between the pictures for wobbling motion
  around 1-axis and 3-axis are consistent with the pictures of components of angular
  momentum shown in Fig.~\ref{fig5}.


\section{Summary}\label{sec5}

  In summary, a simple triaxial rotor is taken as an example to illustrate the picture
  of wobbling motion by exactly solving the TRM Hamiltonian. The obtained energy spectra
  and reduced quadrupole transition probabilities are compared to those approximate analytic
  solutions by HA~\cite{Bohr1975} and HP~\cite{Tanabe1971PLB, Tanabe2006PRC,  Tanabe2008PRC}
  methods to evaluate the accuracy of the analytic methods. It is found that only the low
  lying wobbling bands can be well described by the analytic HA solutions. By taking the higher
  order terms into account, the HP method improves the HA descriptions for the energy spectra.

  The evolution of angular momentum geometry with respect to the spin and the wobbling phonon
  number are investigated. It is confirmed that with the increasing of wobbling phonon number,
  the triaxial rotor changes its rotation along the axis with the largest MOI to
  the axis with the smallest MOI. There is a competitive process among three
  angular momentum components in the separatrix region in which the angular momentum mainly
  aligns along the axis with intermediate MOI. Nevertheless, a specific track that can be used
  to depict the motion of a triaxial rotating nuclei at any spin is found along the
  evolutionary process. This evolutionary track may be regarded as a fingerprint of the triaxial
  rotor. The $K$-distribution further corroborates the evolution picture of the angular momentum.

  For a perspective, it is interesting to investigate the angular momentum geometry of wobbling
  motion for odd-$A$ nuclei, in particular the angular momentum geometry of longitudinal and
  transverse wobblers~\cite{Frauendorf2014PRC}.

\section*{Acknowledgements}

The authors are grateful to Prof. J. Meng and Prof. S. Q. Zhang for fruitful
discussions and critical reading of the manuscript. The work of W. X. Shi
was supported in part by the President's Undergraduate Research Fellowship (PURF),
Peking University. This work was partly supported by the Major State 973 Program of
China (Grant No. 2013CB834400), the National Natural Science Foundation of China
(Grants No. 11175002, No. 11335002, No. 11375015, No. 11345004, No. 11461141002),
the National Fund for Fostering Talents of Basic Science (NFFTBS) (Grant No. J1103206),
the Research Fund for the Doctoral Program of Higher Education (Grant No. 20110001110087).




\end{CJK}

\end{document}